# Next generation of IEEE 802.11 by minimizing the current problems

U.H.D. Thinura Nethpiya Ariyaratne
{it19198200@my.sliit.lk}
Undergraduate, Faculty of Computing, Sri Lanka Institute of Information Technology, Malabe, Sri Lanka

**Abstract**

This main objective of this research is to analyze the common issues of the current and past IEEE 802.11 versions (which is also called as Wi-Fi) that have risen with the increase of internet users and their demands. While this study is conducted with the help of several studies taken into analysis, it provides the solutions that would be able to come out with next generation of IEEE 802.11 in order to minimize the limitations of the current technologies and also to meet the demands of the customers.

**Introduction**

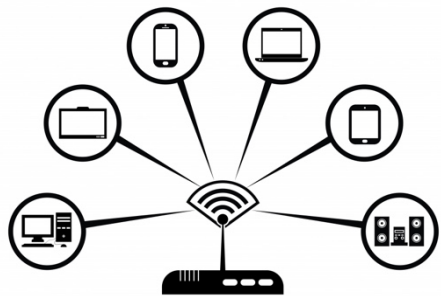

*Figure 2 : Wi-Fi enabled devices connected to a wireless AP*

IEEE 802.11 which is mostly known as Wi-Fi, is a set of rules and regulations created by the Institute of Electrical and Electronics Engineers (IEEE) for wireless local area networks (WLANs). [1]

Wi-Fi wireless networking has become an essential portion of our ordinary routine as it is built across all cellphones which are being one of the essential components of our lives today, providing for minimal cost network connectivity. IEEE 802.11 is also used by desktops, laptops, tablet devices, cameras, and a wide range of other equipment to obtain network connectivity as shown in Figure 1. Moreover, smart homes have become a widespread focus these days where many networking devices are required for these models which use Wi-Fi for communication and automation within the houses. Basically, in homes and offices, local area networks are using Wi-Fi majorly for communication along with Ethernet. Because of these things Wi-Fi has become the major carrier of data. So, most of the public places including offices, shopping centers, coffee shops, airports, and also homes will have Wi-Fi access readily available by using Wi-Fi access points or DSL/Ethernet routers.

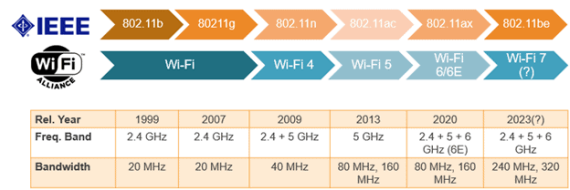

*Figure 1 : Variants of IEEE 802.11*

As many manufacturers are creating Wi-Fi-enabled devices, IEEE created a common standard named 'IEEE 802.11' for Wi-Fi. So, all these devices will have the same platform that enables them to interconnect with each other. IEEE 802.11 will be the base standard while there are also some variants like 802.11n, 802.11ac, 802.11ax of Wi-Fi that have been created as shown in Figure 2.

In order to keep up with the ever-increasing demands for higher bandwidth as well as faster speeds, among other things, these types of newer variations are developed. For example, we can take Gigabit Wi-Fi that is already commonly utilized during these days.

**What problems do we face now?**

**Increased Traffic –**

As Wi-Fi has become very popular among all the people in the world, people tend to use Wi-Fi at each and every place. It can be at the office, at home, in public places, or anywhere because of the portability and ease of use of this technology. Another reason for people using Wi-Fi will be because these devices are very cheaper, and the devices are using unlicensed bands which will usually be 2.4 GHz and 5 GHz. Also, Wi-Fi is a swiftly evolving technology so when each person is using more than one Wi-Fi-enabled device, that will create a collection of thousands and thousands of Wi-Fi devices in a certain area. Due to these reasons, the unlicensed band will become congested. Cisco's *Visual Networking Index* addresses this reason, "it is expected that there will be over 11.6 billion mobile devices producing and consuming a total of 30.6 exabytes of monthly traffic in 2020". [2]

**Lower Bandwidth –**

Network traffic majorly includes video traffic as it keeps growing day by day from 480p to 720p HD and then 4K, 8K, and so on using many bandwidths of data. Other than video traffic, augmented reality (AR), virtual reality (VR), gaming, cloud services, and virtual office work traffic can be taken which uses the bandwidth mostly. [2]

**Lower Reliability –**

Most of these services should not get at least a small delay in sending packets through the network and also there should not be any packet loss, the reliability should be nearly 99%.

These factors will be some of the issues that current Wi-Fi technologies might be having.

Due to these problems, people might tend to use Ethernet technologies rather than using Wi-Fi as it will minimize these issues. [2]

**How can we minimize these issues?**

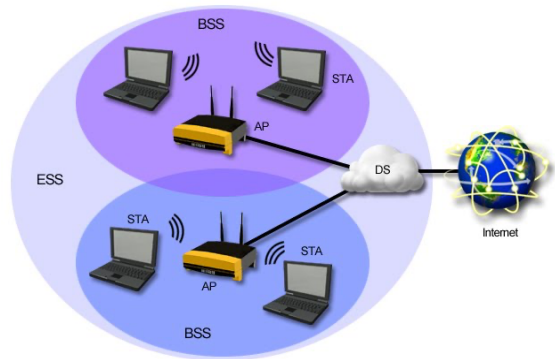

*Figure 3 : Station (STA) in a network*

In the Wi-Fi world, an STA which is also called a 'Station' is a fixed or portable device that is capable of using the IEEE 802.11 protocols. As shown in Figure 3, a laptop computer can be taken as an STA. Most of the STAs support dual-radio stations and access points (APs) these days are supporting triple-band as of now because 6 GHz has also come into action other than 2.4 and 5 GHz. So, the main mission of the next generation IEEE 802.11 is to efficiently use these tri bands and the channels which will increase the speed and customer satisfaction as well. [3]

**Data aggregation in Multiple bands –**

In order to escalate the throughput of the next generation IEEE 802.11 technology, we can combine both the 5 GHz and 6 GHz spectrums not only for transmission of data but also for signal reception. After aggregating both of the bands, any Wi-Fi device that is planning to start a communication need to synchronize the transmit opportunity (TXOP) which is a feature in the MAC layer that is used in WLANs. TXOP is a feature that is used to limit the time duration allocated for a certain STA to send frames after it has been given the opportunity to communicate in the medium. This ensures that there will be no collisions in the transmission media which will be really important when video or voice data of a high throughput is transmitted. So, using these techniques will ensure that there will be a smoother communication between the two spectrums, even though they are aggregated. [4]

**Data transmission and signal reception concurrently through unlike bands or channels –**

By offering serial and concurrent upload or download activity over several independent bands or channels, seems to have the ability to reduce transmission delays as well as to increase throughput. This property is also known as a multi-band or multi-channel full-duplex. This can also be included in the next generation IEEE 802.11 version. But, to reduce the interference that will occur during uplink-to-downlink and downlink-to-uplink communications, we can include a basic segregation between the uplink and downlink channels. [4]

**Data transmission and signal reception concurrently using like bands or channels –**

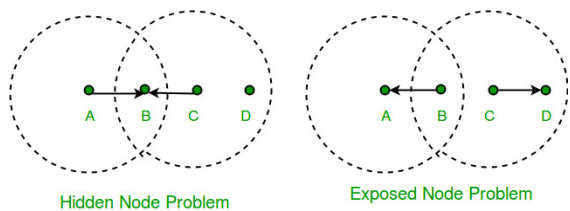

*Figure 5 : Hidden node in a network*

To minimize the latency, maximize the throughput for each STA, develop the collision detection mechanisms, and reduce the hidden node issue in Wi-Fi within a very crowded basic service set (BSS), we can use full-duplex communication methods in bands or channels [4]. A hidden node in a Wi-Fi network is an issue that is occurred when a certain device in the network is able to communicate with a wireless AP but cannot straightaway communicate with the rest of the devices that are connected to that certain wireless AP for network communication. This is also shown in Figure 5. [5]

**Separation of the control plane and the data plane in Wi-Fi –**

Most of the time in the current Wi-Fi technologies, one channel is used for both data transmission and to get details about the control information like the STA buffer condition. This will create a huge delay, huge overheads in data transmission and also will result in high throughput.

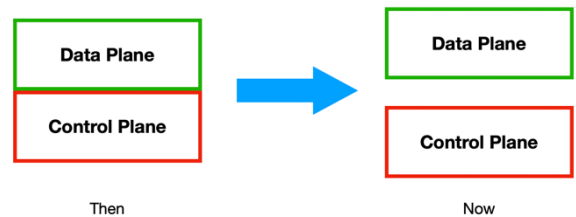

*Figure 4 : Separation of data and control planes*

But if we can introduce multiple separate bands or channels where one for data transmission purposes and then another channel to get updates of control information on a regular and consistent basis, we can minimize the above-mentioned issues. The plane separation is shown in Figure 4. Furthermore, as the new IEEE 802.11 will be a multi-band or multi-channel with all being a full duplex, there is a higher chance of increasing the quality of the service by separating the data transmission and network management levels by implementing this mechanism. [4]

**Implementing a Distributed Multiple-Input Multiple-Output (D-MIMO) –**

MIMO is a wireless transmission antenna system comprising an array of antennas at both the producer as well as the recipient. [6] In order to reduce faults, maximize data speed, and increase radio transmission bandwidth whilst allowing data to be transferred across several transmission routes at the very same moment, these transmitters at either side of the communication media get integrated. A Distributed MIMO or a D-MIMO will have multiple sets of MIMO interconnected together to form an array of virtual antenna. This can be also referred to as a virtual MIMO system, yet another approach towards eliminating the basic foundational limitations of MIMO itself. [7]

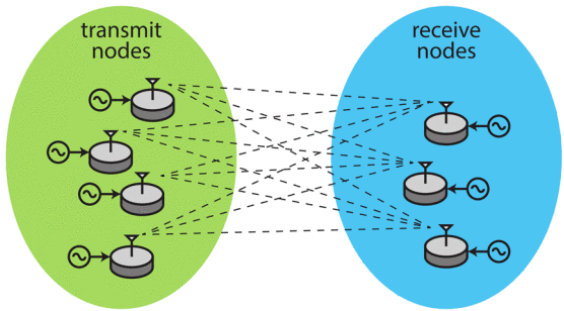

*Figure 6 : A model of distributed a MIMO (D-MIMO) where each sender (transmitter) and receiver nodes are having their own oscillator.*

As shown in Figure 6, by using a consolidated array of antennas this kind of distributed MIMO system could perhaps recreate each and every MIMO communication strategy. It will provide an enhanced transmit/receive antenna directivity. Where directivity is referred to as the measuring scale of the concentration of a certain antenna's model of radiation towards a direction that is measure in dB (Decibels) [8]. If there is a higher level of directivity, that means the antenna directs a highly concentrated beam which also means that the beam is capable of moving a greater distance. Minimizing the interference, elevated degrees of freedom and spectral efficiency, enhanced spatial diversity will be some other advantages of this new edition of MIMO that surpasses the conventional MIMO.

Furthermore, the most effective way of achieving genuinely enormous MIMO systems would be by creating a D-MIMO solution that would be very scalable without the consideration of the number of nodes that would participate.

There is a need to address the base limitations that were found in synchronization, monitoring the autonomous oscillator and motion interactions, and also building an extensible, reduced latency methodology for a strong distributed MIMO system to be built upon the older MIMO system, so that the migration from traditional MIMO network with poor coordination omnidirectional transmitters to an intelligent network of organized D-MIMO endpoints will be a smooth transition.

**Developing the Retransmission Protocol of the links -**

When a certain packet is transmitted to the receiver and an acknowledgment (ACK) packet is not received back to the sender or if there are any problems occurred when decoding the packet at the receiver's end, the same error packet will be retransmitted. This is because, in the current Wi-Fi technologies, they are considering only the MAC Protocol Data Units or sometimes referred to as MPDUs which uses a mechanism called the Automatic Repeat Request (ARQ). In here the error MPDU will be discarded by the receiver as soon as it receives at their end, without waiting for the sender's retransmission of the remaining missing parts which can be combined to the initial packet has already received at the receiver's end. [7]

As this is a major issue when considering the reliability and the latency of data transmission, there is a need of a new ARQ mechanism. Hybrid Automatic Repeat Request (HARQ) can be introduced to overcome these issues as it will not request to retransmit or discard the former error packets, instead, it will request the former problematic packet with the new packet concurrently and combine the soft bits of the error packet so that there will be no issues while decoding it. Something to highlight is that in here this mechanism is capable of receiving the problematic packet from the receiver with another new packet that was queued to be transmitted. So, this will increase the reliability and reduce the latency which will positively impact the bandwidth and the speed of the network.

In the mobile communications sector, this procedure is already carried out. When talking about some issues with HARQ, compared to ARQ it needs additional processing resources and also a higher amount of storage is required for collecting previous messages to the process of soft combining the problematic packets. But if some more researching activities are conducting on this topic, it would be possible to overcome these issues and provide the high performance and reliable HARQ that is required for IEEE 802.11. The differences between ARQ and HARQ are further shown in Figure 7.

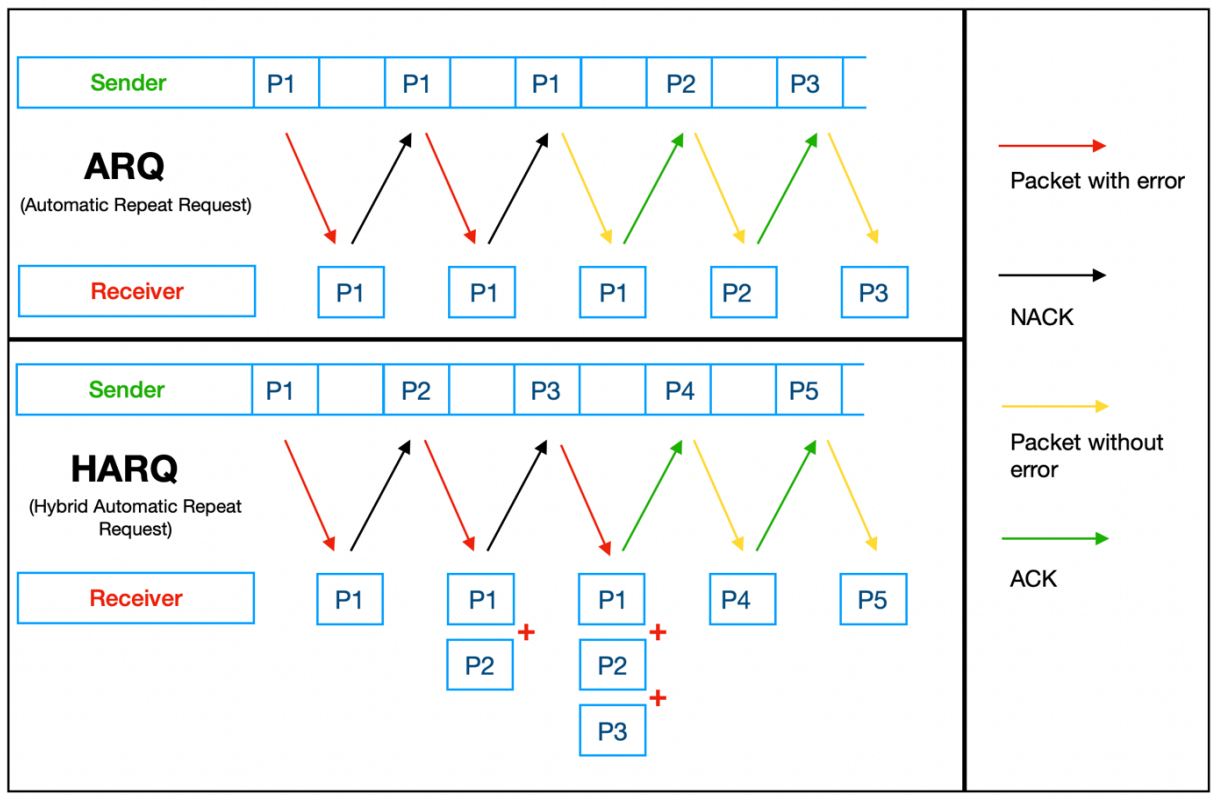

*Figure 7 : Comparison of packet transmission in ARQ and HARQ mechanisms*

## Conclusion

With the background of how Wi-Fi has come into being, the ever-increasing users and their ever-increasing demand for better internet is one of the most known issues globally. Stating the current issues faced like traffic, low bandwidth and reliability, the next generation IEEE 802.11 provides the most optimal solutions to minimize the limitations of the older generation. While the adoption to the new generation will take time, by then more research aiding and backing the 802.11ax will come into being.